\documentstyle[preprint,aps]{revtex}
\begin{document}
\title{Coherent Control of Multiphoton Transitions with Femtosecond pulse 
shaping}
\author{S. Abbas Hosseini and Debabrata Goswami} 
\address{
Tata Institute of Fundamental Research, Homi Bhabha Road,
Mumbai 400 005, India.
}
\date{\today}
\maketitle
\begin{abstract}
We explore the effects of ultrafast shaped pulses for two-level systems that 
do not have a single photon resonance by developing a multiphoton 
density-matrix approach.  We take advantage of the fact that the dynamics of 
the intermediate virtual states are absent within our laser pulse 
timescales.  Under these conditions, the multiphoton results  are similar to 
the single photon and that it is possible to extend the single photon 
coherent control ideas to develop multiphoton coherent control.  
\end{abstract}


\section{Introduction} 

Use of optimally shaped pulses to guide the time evolution of a system and 
thereby control its future is an active field of research in recent 
years \cite{ct1}-\cite{ct16}.  Such developments have been spurred by technological 
breakthroughs permitting arbitrarily amplitude modulated laser pulses with 
20-30 fs resolution and pulse energies ranging to almost hundred 
microjoules--either in the time domain or in the frequency domain.  In most 
practical cases, computer optimizations are used to generate the useful 
shapes\cite{ct1}-\cite{ct7}, since even approximate analytical solutions exist only 
for very specialized cases\cite{ct7}-\cite{ct12}.  Such computer simulations have 
resulted in generating quite complicated theoretical waveforms that can 
break strong bonds\cite{ct1}-\cite{ct4}, localize excitation\cite{ct13}.  Most of 
these interesting calculations involve intense pulses, which do not operate 
in the linear response regime.  Actual photochemical processes with such 
intense pulses that operate beyond the linear response region often involve 
multiphoton effects.  Unfortunately, multiphoton interactions typically 
induce additional complications and have not yet been explored much 
theoretically for coherent control purposes.  In fact, most models for 
coherent control deal with light-matter interaction at the single-photon 
level.  However, some recent experiments show that they can even simplify 
quantum interference effects; e.g., how Cs atoms can be made to absorb or 
not absorb light with non-resonant two-photon excitation with shaped optical 
pulses\cite{ct14,ct15}.  The experimental results have been treated with a 
perturbation model that works under the resonant condition.  However, a more 
complete theoretical treatment of multiphoton interactions for developing 
multiphoton coherent control is quite complex and is far from complete.  In 
fact, the lack of such a theoretical basis is also evident from the fact 
that in the classic demonstration of control of multiphoton ionization 
process, an experimentally optimized feedback pulse shaping was found to 
provide the best-desired yield\cite{ct16}.  In the present work, we develop 
a density matrix approach to multiphoton processes that do not have any 
lower-order process and demonstrate that can also explain the off-resonance 
behavior.  

We present results, which show that this would be a promising approach.  We 
first apply the approach to the two-photon scenario in a simple two-level 
system (e.g., any narrow, single-photon transition line that is only 
multiphoton allowed).  We then generalize the results to the case where 
only one N-photon (N$\ge$2, which is multiphoton) transition is possible 
and none of the (N-1) photon transition can exist.  Under these conditions, 
we show that most of the waveforms produce the same results as the single 
photon case\cite{ct18}.  With care, therefore, we predict that it will be 
possible to extend some of the single photon coherent control ideas to 
develop multiphoton coherent control.  We explore the various 
frequency-swept pulses into the multiphoton domain, which have been 
previously shown to be successful in inducing robust inversions under 
single-photon adiabatic conditions.  We also investigate the case of phase 
modulated overlapped Gaussian pulses for two-photon transition (in the 
spirit of a ``dark'' pulse of Meshulach and Silberberg, which they defined 
as ``a single burst of optical field'' that does not produce any net 
population transfer\cite{ct15}).  We show that the two-photon dark pulses, 
which are a result of smoothly varying phase modulation, can be explained by 
invoking the well-established concept of single photon adiabatic rapid 
passage (ARP)\cite{ct17,ct18} to the multiphoton framework.  In fact, the 
ARP explanation allows us to generalize the results to the N-photon case and 
show that such dark pulses are a result of the Stark shifting of the 
resonant N${}^{th}$ photon transition.  The extension of the concept of ARP 
into the multiphoton domain has very important consequences in generating 
inherent robust processes.

\section{Formalism}

The simplest model describing a molecular system is an isolated two-level 
system or ensemble without relaxation or inhomogeneities. This simple model 
often turns out to be a very practical model for most systems interacting 
with the femtosecond laser pulses as the magnitude of the relaxation 
processes are immensely large as compared to the femtosecond interaction 
time. Let us consider a linearly polarized pulse is being applied to the 
$|$1$>$$\rightarrow$$|$2$>$ transition, where $|$1$>$ and $|$2$>$ represent the ground and excited 
eigenlevels, respectively, of the field-free Hamiltonian. In case of single 
photon interactions (Fig. 1a), the total laboratory-frame Hamiltonian for 
such two-level system under the effect of an applied laser field, $E\left( 
{t} \right) = \varepsilon \left( {t} \right)e^{i\left[ {\omega \left( {t} 
\right)t + \phi \left( {t} \right)} \right]} = \varepsilon \left( {t} 
\right)e^{i\left[ {\omega + \dot {\phi} \left( {t} \right)} \right]t} $ can 
be written as\cite{ct10,ct17}:

\begin{equation}
H = \left( {{\begin{array}{*{20}c} {E_{1}}  \hfill & {V_{12}}  \hfill \\
 {V_{21}}  \hfill & {E_{2}}  \hfill \\
\end{array}} } \right) = \left( {{\begin{array}{*{20}c}
 {\hbar \omega _{1}}  \hfill & {\mu .E} \hfill \\
 {\mu .E^{\ast} }  \hfill & {\hbar \omega _{2}}  \hfill \\
\end{array}} } \right) = \hbar \left( {{\begin{array}{*{20}c}
 { - \frac{{\omega _{R}} }{{2}}} \hfill & {\frac{{\mu .\varepsilon} }{{\hbar 
}}e^{i\left( {\omega \,t + \phi}  \right)}}  \hfill \\
 {\frac{{\mu .\varepsilon ^{\ast} } }{{\hbar} }e^{ - i\left( {\omega \,t + 
\phi}  \right)}}  \hfill & {\frac{{\omega _{R}} }{{2}}} \hfill \\
\end{array}} } \right)
\end{equation}

\noindent
where $\omega _{R} = \omega _{2} - \omega _{1} $ is resonance frequency, 
{\it V}$_{12}$ and {\it V}$_{21}$ are the negative interaction potentials 
and $h\omega _{1} ,{\kern 1pt} \,h\omega _{2} $ are the energies of ground 
({\it E}$_{1}$) and excited state ({\it E}$_{2}$) respectively, and $\mu $ 
is the transition dipole moment of the $|$1$>$$\rightarrow$$|$2$>$ transition. In analogy 
to this single photon interaction as given in Eqn. (1), the interaction 
potential under the effect of an applied laser field, in two-photon 
absorption case (Fig. 1b) can be written as:

\begin{equation}
V\left( {t} \right) = \mu _{1m} \varepsilon \left( {t} \right)\,e^{i\left( 
{\omega \,t + \phi \left( {t} \right)} \right)} \,\left| {1} \right\rangle 
\left\langle {m} \right|\,\mu _{m2} \varepsilon \left( {t} 
\right)\,e^{i\left( {\omega \,t + \phi \left( {t} \right)} \right)} \,\left| 
{m} \right\rangle \left\langle {2} \right| + c.c.
\end{equation}

\noindent
where {\it m} is the virtual state. Let us, for simplicity, take the 
transition dipole moment between the ground state to the virtual sate to be 
equal to the transition dipole moment between the virtual state and excited 
state $\left( {\mu _{1m} = \mu _{m2} = \mu}  \right)$. In fact, we have 
verified in our simulations that the trend of the results is preserved even 
when we relax this simplification. In any event, for developing the initial 
model, the above said simplification allows us to take $\mu $ as a common 
factor and we can rewrite Eqn. (2) as:

\begin{equation}
V\left( {t} \right) = \left( {\mu \,\varepsilon \left( {t} \right)} 
\right)^{2} \,e^{2i\left( {\omega \,t + \phi \left( {t} \right)} \right)} 
\left| {1} \right\rangle \left\langle {2} \right| + c.c.
\end{equation}

\noindent
since for normalized states, $ < m\,|m > \, = 1$. Using similar arguments 
for the {\it N}-photon case (Fig. 1c), the interaction potential can be 
written as:

\begin{equation}
V\left( {t} \right) = \left( {\mu \,\varepsilon \left( {t} \right)} 
\right)^{N} \,e^{iN\left( {\omega \,t + \phi \left( {t} \right)} \right)} 
\left| {1} \right\rangle \left\langle {2} \right| + c.c.
\end{equation}

Thus, the total laboratory-frame {\it N}-photon Hamiltonian will be:

\begin{equation}
H = \left( {{\begin{array}{*{20}c} {\hbar \omega _{1}}  \hfill & {\left( {\mu .E} \right)^{N}}  \hfill \\
 {\left( {\mu .E^{\ast} }  \right)^{N}}  \hfill & {\hbar \omega _{2}}  \hfill \\
\end{array}} } \right) = \hbar \left( {{\begin{array}{*{20}c} { - \frac{{\omega _{R}} }{{2}}} \hfill & {\frac{{\left( {\mu .\varepsilon}  
\right)^{N}} }{{\hbar} }e^{iN\left( {\omega \,t + \phi}  \right)}}  \hfill 
\\
 {\frac{{\left( {\mu .\varepsilon ^{\ast} }  \right)^{N}} }{{\hbar} }e^{ - 
iN\left( {\omega \,t + \phi}  \right)}}  \hfill & {\frac{{\omega _{R} 
}}{{2}}} \hfill \\
\end{array}} } \right)
\end{equation}

The virtual levels for the two-photon (or N-photon) case can exist anywhere 
within the bandwidth $\Delta \omega $ of the applied laser pulse (Fig. 1) 
and the individual virtual state dynamics is of no consequence. 

In analogy to the single photon case\cite{ct11,ct12}, there are two different 
ways to transform the elements of the above laboratory frame N-photon 
Hamiltonian (Eqn. (5)) into a rotating frame of reference. Any 
time-dependent transformation function {\it T} can be applied on both sides 
of the Schrodinger equation as follows:

\begin{equation}
\begin{array}{l}
 T\left( {i\hbar \frac{{\partial} }{{\partial t}}\Psi = H\Psi}  \right) \\ 
 i\hbar \frac{{\partial} }{{\partial t}}\left( {T\Psi}  \right) - i\hbar 
\frac{{\partial T}}{{\partial t}}\left( {T^{ - 1} T} \right)\Psi = TH\left( 
{T^{ - 1} T} \right)\Psi \\ 
 i\hbar \frac{{\partial} }{{\partial t}}\left( {T\Psi}  \right) = \left[ 
{THT^{ - 1} + i\hbar \frac{{\partial T}}{{\partial t}}T^{ - 1}}  
\right]\left( {T\Psi}  \right) \\ 
 \end{array}
\end{equation}

\noindent
which results the following transformation equation:

\begin{equation}
H^{Transformed} \quad = \quad THT^{-1}  + i\hbar T^{-1} \frac{{\partial T}}{{\partial t}}
\end{equation}

The usual frame of reference would be to rotate at {\it N$\omega $}. This is 
the phase-modulated (PM) frame of reference, which can be derived from the 
Hamiltonian {\it H} of Eqn. (5) by the transformation:

\begin{equation}
T^{PM} = \left( {{\begin{array}{*{20}c}
 {e^{ - iN\frac{{\omega \,t}}{{2}}}}  \hfill & {0} \hfill \\
 {0} \hfill & {e^{iN\frac{{\omega \,t}}{{2}}}}  \hfill \\
\end{array}} } \right)
\end{equation}

Using of Eqn. (7), the transformed Hamiltonian in the PM frame is:

\begin{equation}
H^{PM} = \hbar \left( {{\begin{array}{*{20}c}
 {\Delta}  \hfill & {\frac{{\mu \left( {\varepsilon \left( {t} \right)} 
\right)^{N}} }{{\hbar} }e^{iN\phi} }  \hfill \\
 {\frac{{\mu \left( {\varepsilon ^{ *}  \left( {t} \right)} \right)^{N} 
}}{{\hbar} }e^{ - iN\phi} }  \hfill & {0} \hfill \\
\end{array}} } \right)
\end{equation}

\noindent
under the assumption that the transient dipole moment of the individual 
intermediate virtual states in the multiphon ladder all add up 
constructively to the final state transition dipole moment and can be 
approximated to a constant ($\mu $) over the N-photon electric field 
interaction. This approximation is particularly valid for the case of 
multiphoton interaction with femtosecond pulses where no intermediate 
virtual level dynamics can be observed. Thus, we define multiphoton Rabi 
Frequency, as the complex conjugate pairs: $\Omega $(t)=$\mu $.($\varepsilon 
$(t))$^{N}$/$\hbar $ and $\Omega ^{*}$(t)=$\mu $.($\varepsilon 
^{*}$(t))$^{N}$/$\hbar $, and the time-independent multiphoton detuning as: 
$\Delta = \omega _{R} - N\omega $ (Fig. 1c). However, in order to 
investigate the off-resonance behavior of continuously modulated pulses, in 
the single photon case, it is useful to perform an alternate rotating-frame 
transformation to a frequency modulated (FM) frame with the transformation 
function:

\begin{equation}
T^{FM} = \left( {{\begin{array}{*{20}c}
 {e^{ - iN\frac{{\omega \,t + \phi} }{{2}}}}  \hfill & {0} \hfill \\
 {0} \hfill & {e^{iN\frac{{\omega \,t + \phi} }{{2}}}}  \hfill \\
\end{array}} } \right)
\end{equation}

\noindent
to transform the N-photon laboratory Hamiltonian in Eqn. (5) to the FM 
frame as:

\begin{equation}
H^{FM} = \hbar \left( {{\begin{array}{*{20}c}
{\Delta + N\dot {\phi} \left( {t} \right)} \hfill & {\frac{{\mu .\left( {\varepsilon \left({t}\right)} \right)^{N}} }{{\hbar} }} \hfill \\ {\frac{{\mu .\left( {\varepsilon \ast \left( {t} \right)} \right)^{N} }}{{\hbar} }} \hfill & {0} \hfill \\\end{array}} } \right) = \hbar \left( {{\begin{array}{*{20}c} {\Delta + N\dot {\phi} \left( {t} \right)}  \hfill & {\Omega}  \hfill \\ {\Omega ^{\ast} }  \hfill & {0} \hfill \\
\end{array}} } \right)
\end{equation}

The time derivative of the phase function $\dot {\phi} \left( {t} \right)$, 
{\it i.e.}, frequency modulation, appears as an additional resonance offset 
over and above the time-independent detuning $\Delta $, while the direction 
of the field in the orthogonal plane remains fixed. The time evolution of 
the unrelaxed two-level system can then be evaluated by integrating the 
Liouville equation\cite{ct10,ct17}:

\begin{equation}
\frac{{d\rho \left( {t} \right)}}{{dt}} = \frac{{i}}{{\hbar} }\left[ {\rho 
\left( {t} \right),H^{FM} \left( {t} \right)} \right]
\end{equation}

\noindent
where $\rho $({\it t}) is a 2$ \times $2 density matrix whose diagonal 
elements represent populations in the ground and excited states and 
off-diagonal elements represent coherent superposition of states. This 
approach has been very successful in solving many single-photon inversion 
processes for arbitrarily shaped amplitude and frequency modulated 
pulses\cite{ct12},\cite{ct13}. We have just extended the same arguments to the 
multiphoton case. 

The simulations are performed with a laser pulse that either has (a) a 
Gaussian intensity profile or (b) a hyperbolic secant intensity profile 
which have the following respective forms:

\begin{equation}
{\begin{array}{*{20}c}
 {\left( {a} \right)\;\;\;\;\;\;\;\;\;\;\;\;\;\;\;\;\;\;\;\;\;\;\;I\left( 
{t} \right) = I_{0} exp\left[ { - \,8ln2\,\left( {t/\tau}  \right)^{2}}  
\right]} \hfill \\
 {\;\;\;\;\;\;\;\;\;\;\;\;implies\;\;\;\varepsilon \left( {t} \right) = 
\varepsilon _{0} exp\left[ { - \,4ln2\,\left( {t/\tau}  \right)^{2}}  
\right]} \hfill \\
 {\left( {b} \right)\;\;\;\;\;\;\;\;\;\;\;\;I\left( {t} \right) = I_{0} 
sech^{2} \left[ {\left\{ {2ln\,\left( {2 + \sqrt {3}}  \right)} 
\right\}\left( {t/\tau}  \right)} \right]}  \hfill \\
 {\;\;\;implies\;\;\;\varepsilon \left( {t} \right) = \varepsilon _{0} 
sech\left[ {\left\{ {2ln\,\left( {2 + \sqrt {3}}  \right)} \right\}\left( 
{t/\tau}  \right)} \right]} \hfill \\
\end{array}}
\end{equation}

\noindent
where $\tau ${\it} is the full width at half maximum, and {\it I(t)} is the 
pulse intensity. This is because most of the commercially available pulsed 
laser sources have these intrinsic laser parameters. We choose a range of 
frequency sweeps, such as (c) the linear frequency sweep for the Gaussian 
amplitude, (d) the hyperbolic tangent sweep for the hyperbolic secant 
amplitude, and they have the following respective forms:

\begin{equation}
{\begin{array}{*{20}c}
 {\left( {c} \right)\;\;\;\dot {\phi} \left( {t} \right) = bt\quad \quad 
\quad \quad \quad \quad \quad \quad \quad \quad \quad \quad \quad \quad 
\quad \;} \hfill \\
 {\left( {d} \right)\;\;\;\dot {\phi} \left( {t} \right) = b\left\{ 
{2ln\left( {2 + \sqrt {3}}  \right)} \right\}tanh\left[ {\left\{ {2ln\left( 
{2 + \sqrt {3}}  \right)} \right\}\left( {t/\tau}  \right)} \right]} \hfill 
\\
\end{array}}
\end{equation}

\noindent
where {\it b} is a constant. Such pulses have been shown to invert 
population through ARP in single photon case and so we choose to use these 
particular shapes for the multiphoton case. 

We also use the shaped overlapping Gaussian pulses for a two-photon 
transition similar to the ones used by Meshulach and Silberberg. In their 
case the frequency sweep is given by:

\begin{equation}
\dot {\phi} \left( {t} \right) = \left\{ {{\begin{array}{*{20}c}
 {b\;\,\,\quad \quad t \ge t_{0}}  \hfill \\
 { - b\quad \quad t < t_{0}}  \hfill \\
\end{array}} } \right.
\end{equation}

\noindent
where {\it t}$_{0}${\it} is the midpoint of the pulse. This pulse does not 
satisfy the ARP condition and is quite susceptible to the changes in the 
pulse amplitude profile and our results show this in the next section. 
However, if we instead use smoothly varying linear frequency sweeps, either 
changing monotonically as in Eqn. (14c), or linearly approaching and going 
away from resonance as given by:

\begin{equation}
\dot {\phi} \left( {t} \right) = bt,$ where {\it b} changes sign at {\it 
t}$_{0}
\end{equation}

These pulses satisfy the ARP conditions as explained in the next section. 
Dark pulses given by Eqn (16) are thus quite insensitive to the changes in 
the pulse amplitude profile. We also extend our calculations to the N-photon 
case in a simple two-level type of system that supports only an N$^{th}$ 
photon transition and show how the phase switches effect the population 
cycling. These generalizations would become evident when we examine the 
results based on the ARP extended to multiphoton case. 

\section{Results \& Discussion}

The population evaluation in a simple two level system without relaxation 
for one photon absorption (N=1) is shown in Fig. 2 for the pulse shapes 
given by Eqns. (13) and (14). Excitation exactly on resonance creates a 
complete population inversion when the pulse area (the time integral of the 
Rabi frequency) equals $\pi $. However, the population oscillates between 
the ground and excited state as sine function with respect to the Rabi 
frequency. These oscillations are not desirable in most cases involving real 
atoms or molecules. They are washed out by inhomogeneous broadening, the 
transverse Gaussian profile of the laser, and (in the molecular case) 
different values of {$\it \mu.\varepsilon $}. For a single-photon case, as 
discussed in Ref. \cite{ct18}, frequency modulated pulses can instead produce 
adiabatic inversion, which avoids these complications. A linearly frequency 
swept (chirped) laser pulse can be generated by sweeping from far above 
resonance to far below resonance (blue to red sweeps), or alternatively from 
far below resonance to far above resonance (red to blue sweeps). When the 
frequency sweep is sufficiently slow such that the irradiated system can 
evolve with the applied sweep, the transitions are ``adiabatic''. If this 
adiabatic process is faster than the characteristic relaxation time of the 
system, a smooth population inversion occurs with the evolution of the 
pulse, which is the well-known ARP. 

Let us now extend the effect of such laser pulses (given by Eqns. (13) and 
(14)) to a two-photon (N=2) case as derived in our Hamiltonian of Eqn. (11). 
Fig. 3 shows the plots of the upper state population ($\rho _{22}$) as a 
function of applied Rabi frequency and detuning for two photon absorption 
case in the absence of one photon absorption. We find that the results are 
qualitatively the same as the one-photon absorption. In fact, our 
simulations show that for such a simple case of a two-level system, where 
only an N$^{th}$ photon transition is possible, we can extend our 
single-photon results to the N-photon case. The difference lies in the Rabi 
frequency scaling. Thus, for this simple case as defined here, we are able 
to invoke the concept of ARP for multiphoton interaction. 

We next use the overlapping Gaussian pulses (when the overlap is complete it 
collapses into a single Gaussian) with different phase relationships. Our 
simulation shows that for shaped overlapping Gaussian pulses the excited 
sate population depends on the form of the frequency sweep. In figure 4a, 
for the shaped pulse without sweep the population of excited state 
oscillates symmetrically. For a simple monotonically increasing or 
decreasing sweep around resonance, it behaves like a Guassian pulse with 
linear sweep (Fig. 4b). These results essentially confirm another important 
implication of the adiabatic principle: that the exact amplitude of the 
pulse is not very important under the adiabatic limit. Again, for this 
simple case, we are able to invoke the concept of ARP for multiphoton 
interaction to explain the inversion. 

The phase modulated overlapped Gaussian pulses are of interest since 
Meshulach and Silberberg had experimentally switched the phase of the second 
pulse with respect to the first pulse and demonstrated two-photon excited 
state population modulation. However, the phase switch in their pulse shapes 
was abrupt as given by Eqn. (15), and thus did not satisfy the ARP 
condition. As a result the population transfer with such pulses are very 
heavily dependent on the actual shape of the pulse. Figs. 5 shows that the 
upper state population for two photon absorption in the absence of one 
photon absorption is heavily dependent on the nature of the phase step, the 
intensity and the extent of overlap of the pulses. At some particular phase 
switch, there is no excited-state population, and they called it the dark 
pulse. We show that it is indeed true for a specific overlapped amplitude 
profile and intensity for a given phase switching position. These dark 
pulses, however, are sensitive to the exact nature of the amplitude profile 
and intensity.

If instead we choose a smoothly varying linear frequency sweeping to the 
two-photon resonance and then away from resonance, as given by Eqn. (16), 
the results are quite robust to the exact nature of the amplitude profile 
and intensity (Fig. 6). At detuning zero and for small values of Rabi 
frequency, we have some population in excited state. However, when the 
intensity of applied pulse increases, the excited state population returns 
to zero. In other words, we are sending shaped pulse into the two-level 
system but finally there is no excited-state population. Curiously enough, 
for such pulses, the population is asymmetric about detuning from resonance. 
In fact, Fig. 6 clearly shows that the population transfer occurs at some 
non-zero detuning values at higher Rabi frequencies when it does not have 
any excitation at resonance and behaves as a dark pulse. This result can be 
understood by examining the evolution of the dressed states\cite{ct17}-\cite{ct19} 
with time (Fig. 7). When the effect of the pulse cannot be felt by the 
system at very early or and at very late times with respect to the presence 
of the pulse, each dressed state essentially corresponds to the single bare 
state ($\left| {\alpha}  \right\rangle \to \left| {1} \right\rangle $ and 
$\left| {\beta}  \right\rangle \to \left| {2} \right\rangle $). It is only 
during the pulse that the dressed states change in composition and evolve as 
a linear combination of the two bare states. The proximity of these dressed 
states during the pulse essentially determines the population exchange. The 
higher Rabi frequencies cause a stark shift in the dressed states so that at 
resonance there is no population exchange. Under such stark shifted 
condition, resonance occurs at some specific non-zero detuning value where 
Rabi oscillations are seen in Fig. 6. These results are completely general 
for a simple case of a two-level system, where only an N$^{th}$ photon 
transition is possible. The phase change of the overlapping Gaussian pulses 
essentially provide an additional parameter to control the population 
evolution of a simple two-level type of system that supports only an 
N$^{th}$ photon transition.

\section{Conclusions}

In this paper, we have explored the effects of ultrafast shaped pulses for 
two-level systems that do not have a single photon resonance by developing a 
multiphoton density-matrix approach. We took advantage of the fact that 
dynamics of the intermediate virtual states are absent in the femtosecond 
timescales, and demonstrated that many multiphoton results can be surprising 
similar to the well-known single photon results. When we extend the ARP to 
the multiphoton condition, robust population inversion and dark pulses 
become possible that are insensitive to the exact profile of the applied 
electric field. We have shown, therefore, that it is possible to extend the 
single photon coherent control ideas to develop femtosecond multiphoton 
coherent control.

\begin{figure}

\caption{ Schematic of (a) single, (b) two and (c) multiphoton 
processes, respectively. Symbols and notations are defined in text.}

\caption{ Comparison of the excited state population for a single photon 
excitation as a function of Rabi frequency, for (a) a Gaussian pulse (solid 
curve: without any frequency sweep; dashed curve: with linear frequency 
sweep), and (b) a hyperbolic secant pulse (solid curve: without any 
frequency sweep; dashed curve: with hyperbolic tangent frequency sweep).}

\caption{ Excited state population for 2-photon excitation as a function 
of Rabi frequency and detuning for: (a) transform-limited Guassian pulse; 
(b) bandwidth equivalent linearly frequency-swept Gaussian pulse; (c) 
transform-limited hyperbolic secant pulse; and (d) hyperbolic secant pulse 
with hyperbolic tangent frequency sweep. }

\caption{ (a) Excited state population for 2-photon excitation as a 
function of Rabi frequency and detuning for Shaped overlapped Gaussian pulse 
without sweep. (b) Excited state population for 2-photon excitation as a 
function of Rabi frequency and detuning for shaped overlapped Gaussian pulse 
with a monotonically increasing linear sweep. }

\caption{ Excited state population for 2-photon excitation as a function 
of phase step position (i.e., detuning) normalized to the pulse FWHM, $\tau 
$, for two different Rabi frequencies in the case of pulses with phase steps 
as given by Eqn. 15. The results are heavily subjective to the choice of 
parameters (as we show for the two Rabi frequencies used in this Fig. that 
differ by less than 5\%), and are thus non-adiabatic, as discussed in the 
text. }

\caption{ Excited state population for 2-photon excitation as a function 
of Rabi frequency and detuning for shaped overlapped Gaussian pulse with a 
sweep linearly approaching and going away from resonance as given by Eqn. 
16. A contour plot (b) is shown for the 3-D surface plot (a) to better 
represent that the population exchange occurs at some detuned position for 
high Rabi frequencies. }

\caption{ Energies of the two dressed states evolving with time for the 
shaped Gaussian pulse whose population evolution is shown in Fig. 6 at a 
high Rabi frequency for (a) no net population transfer at resonance, (b) the 
Stark-shifted frequency (detuned from resonance on one direction) where the 
Rabi oscillations occur, (c) the Stark-shifted frequency equally detuned 
from resonance to the other side where no Rabi oscillations occur. }
\end{figure}

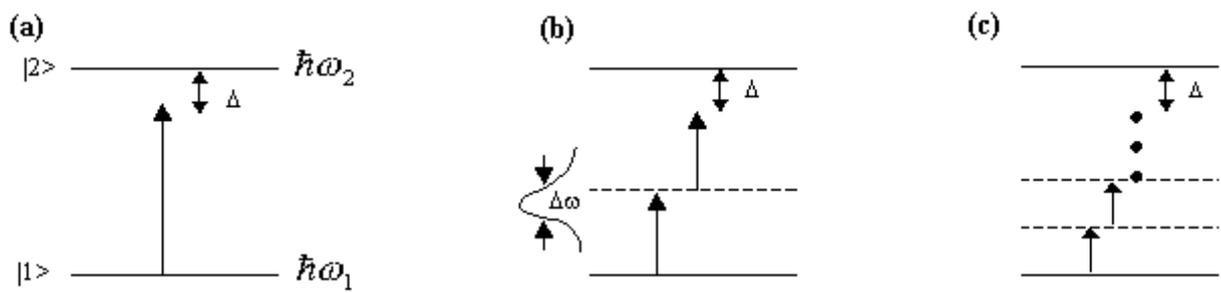
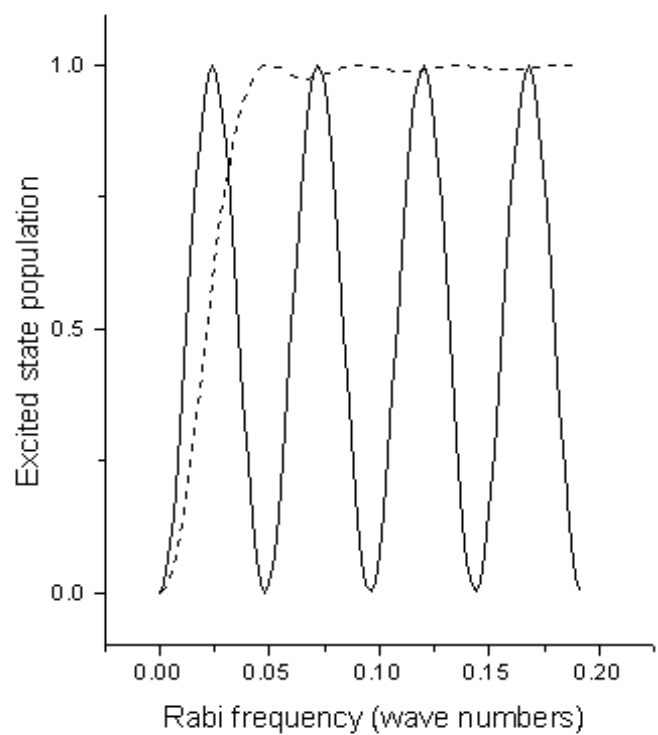
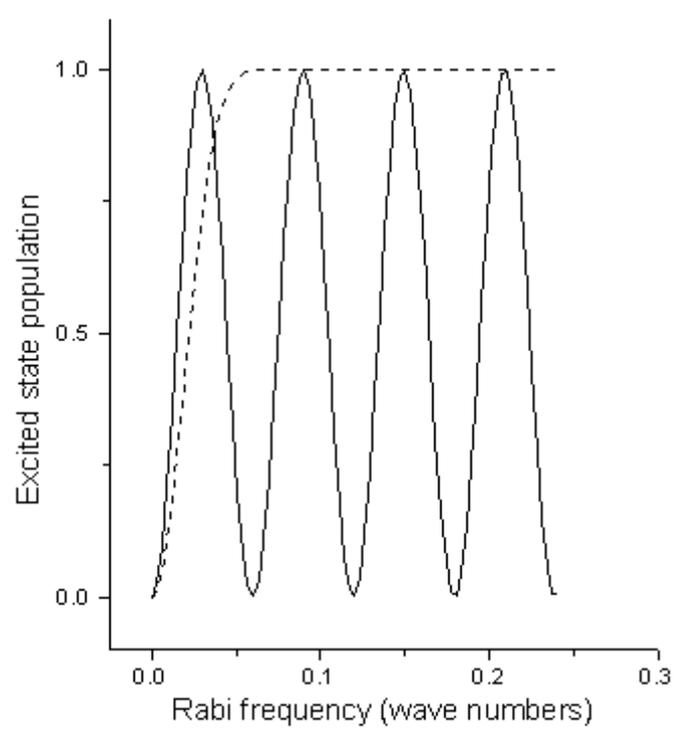
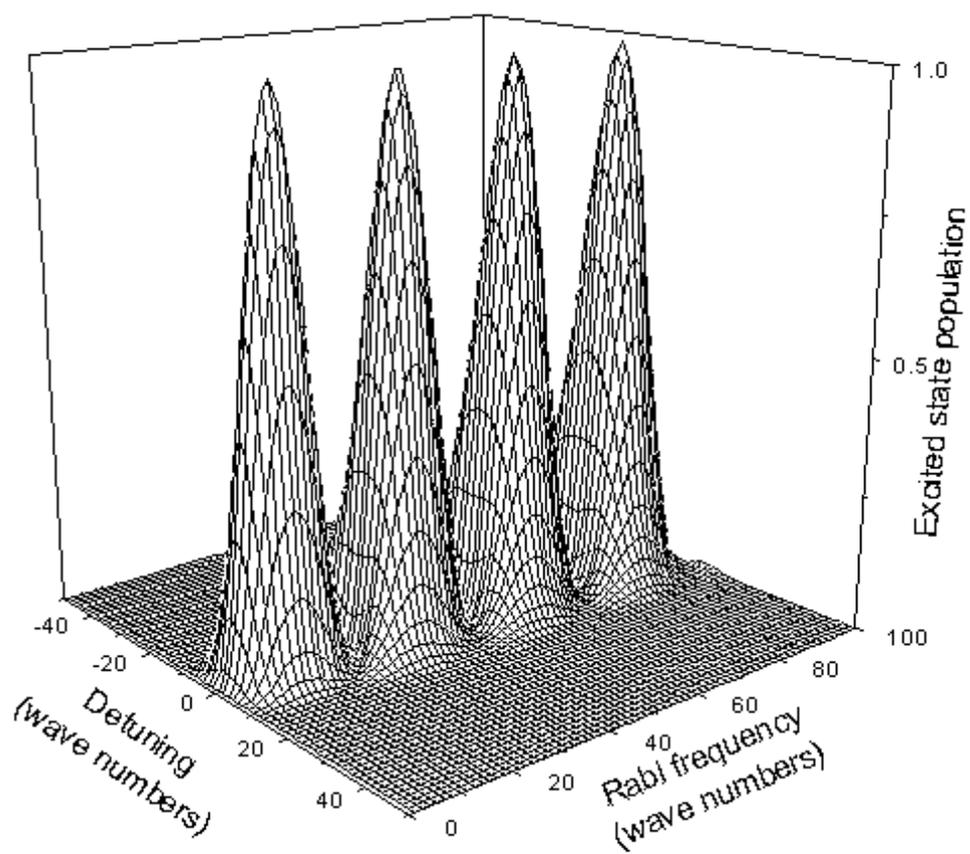
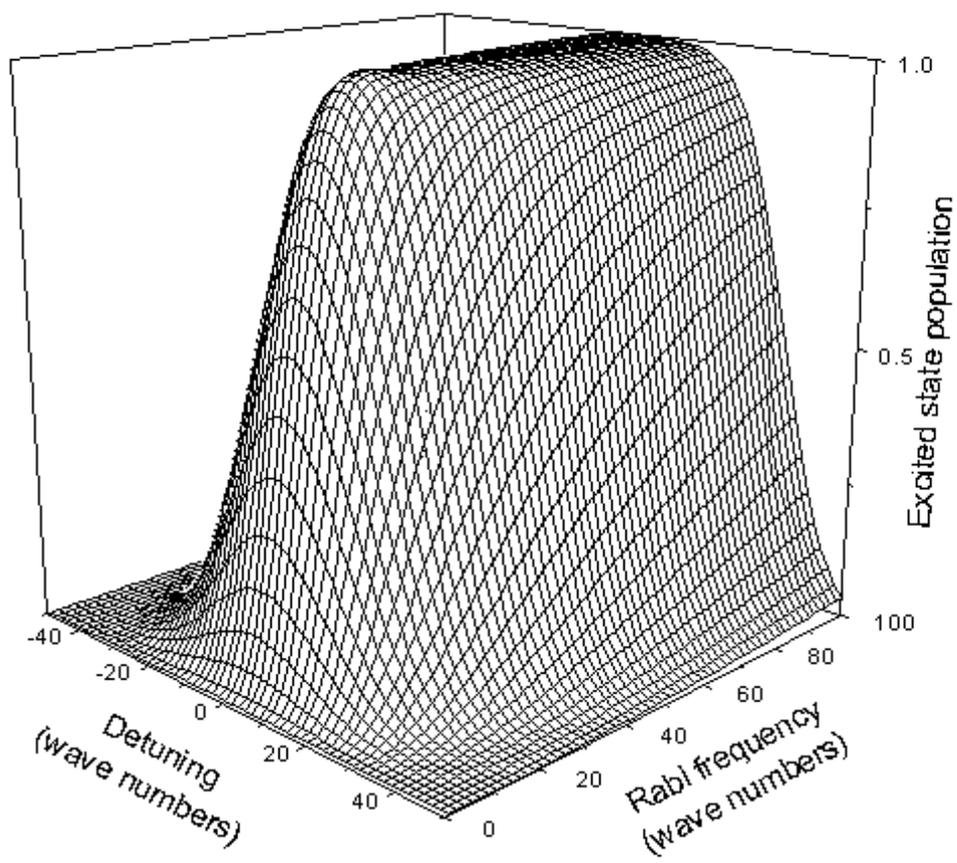
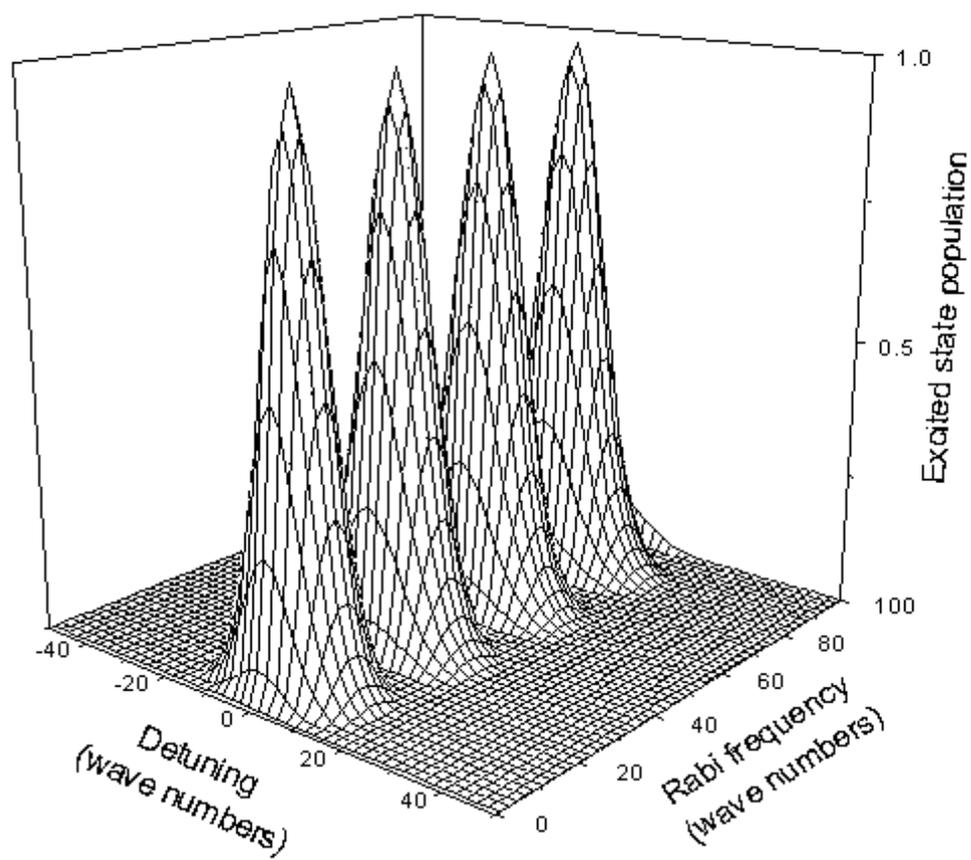
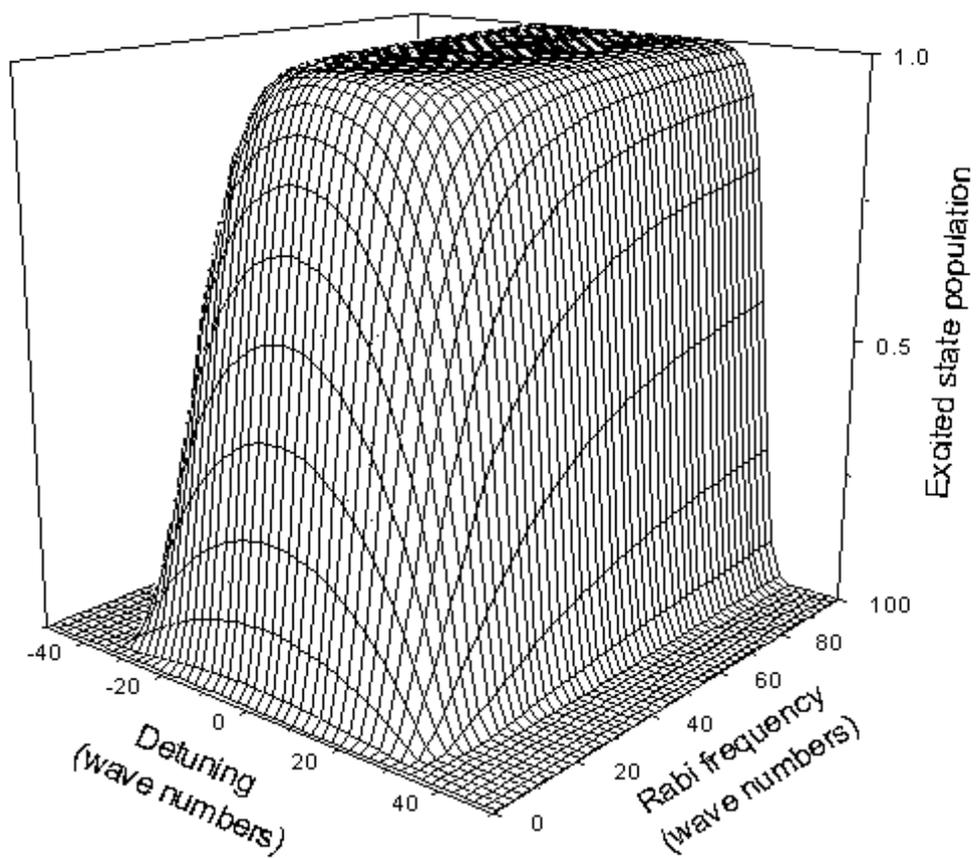
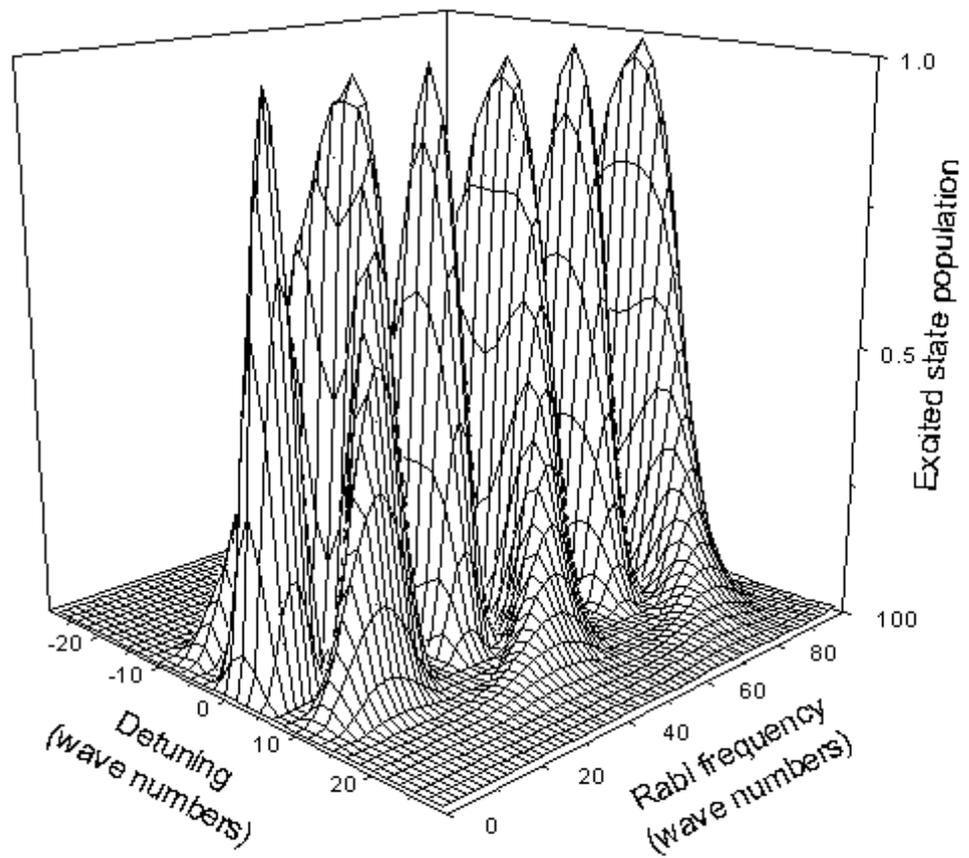
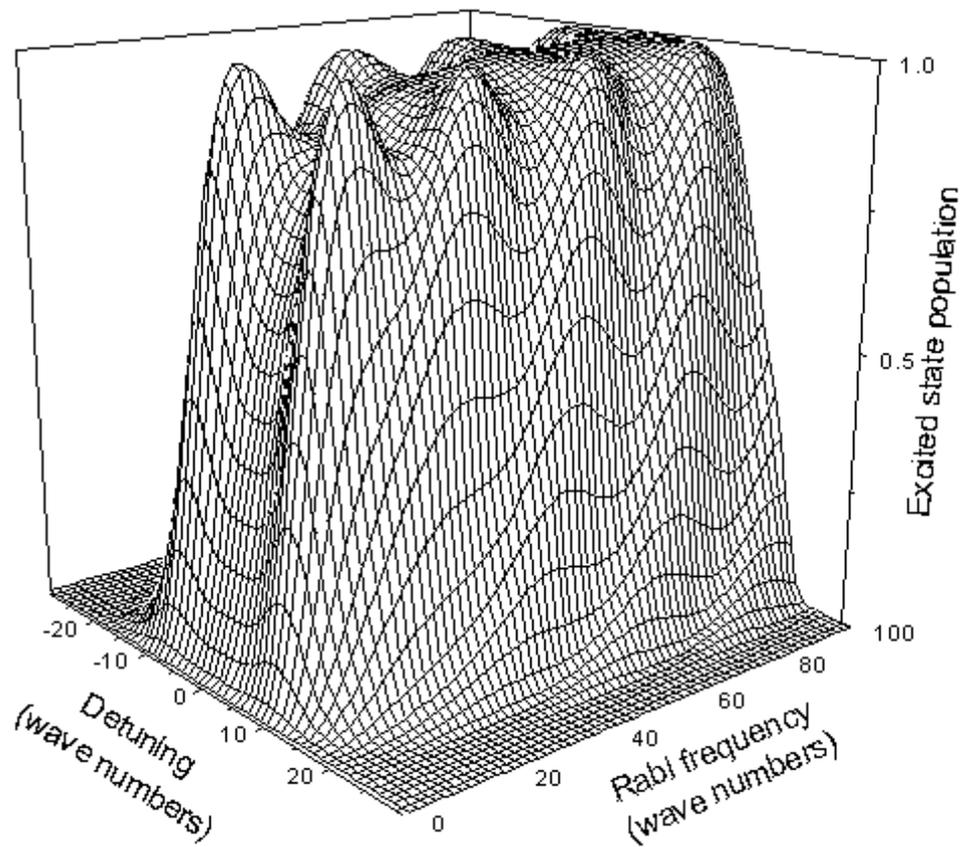
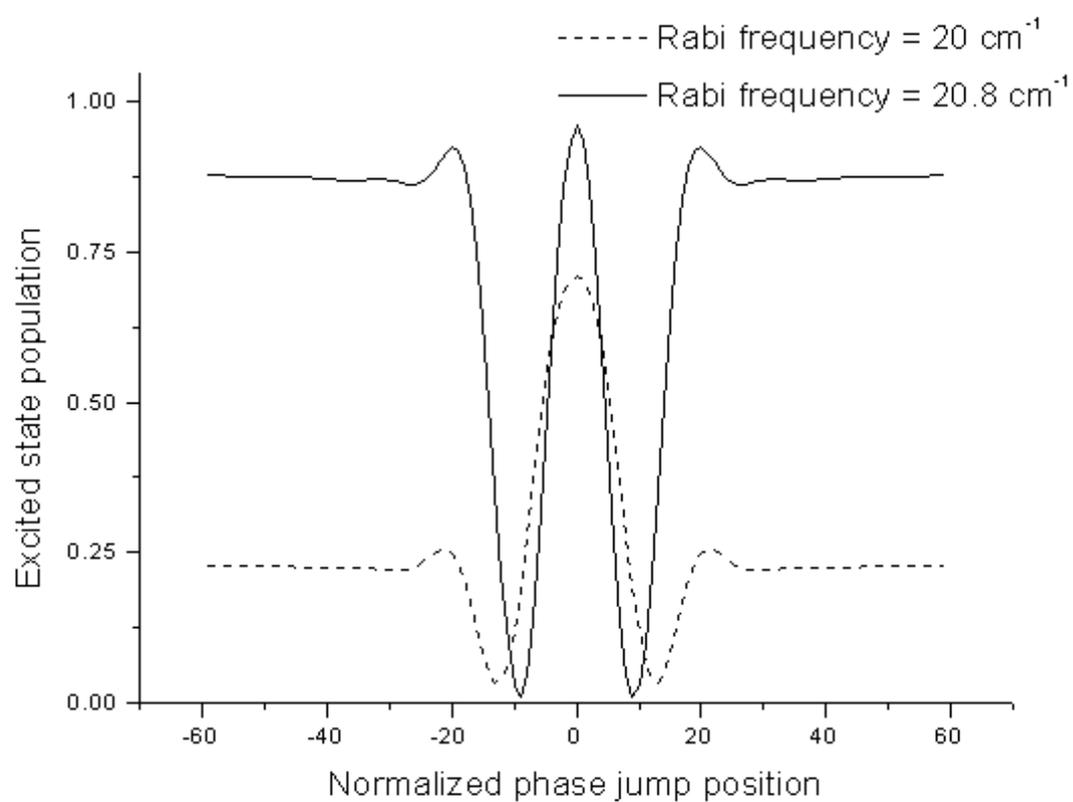
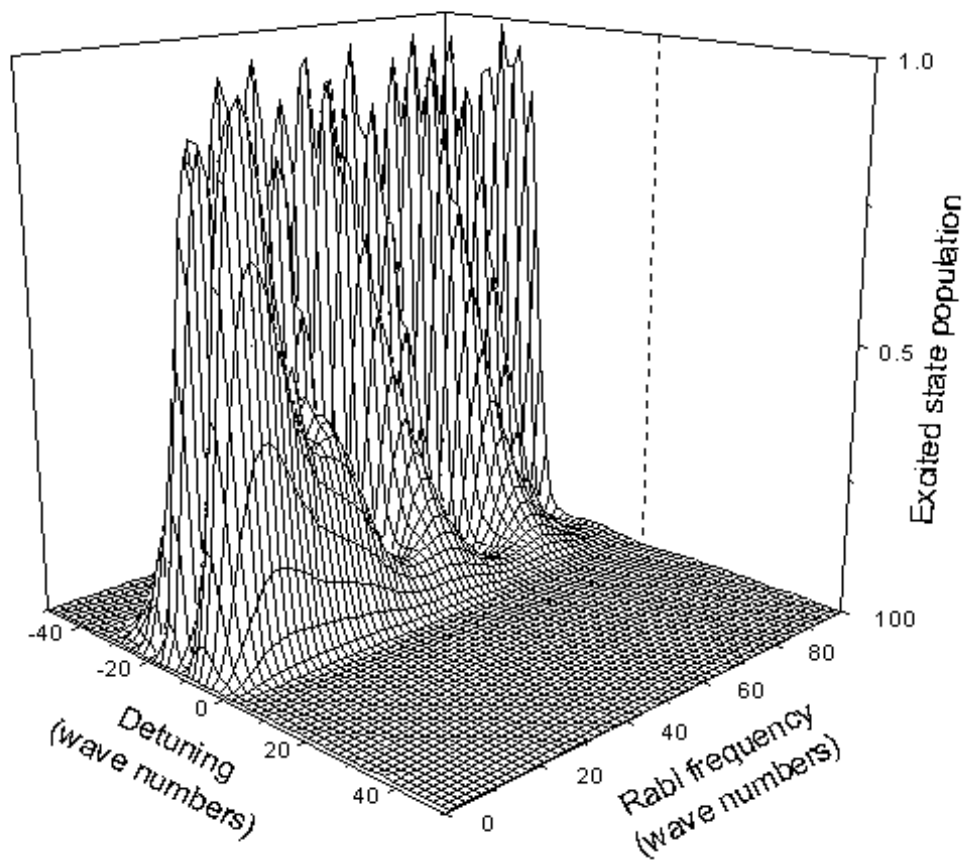
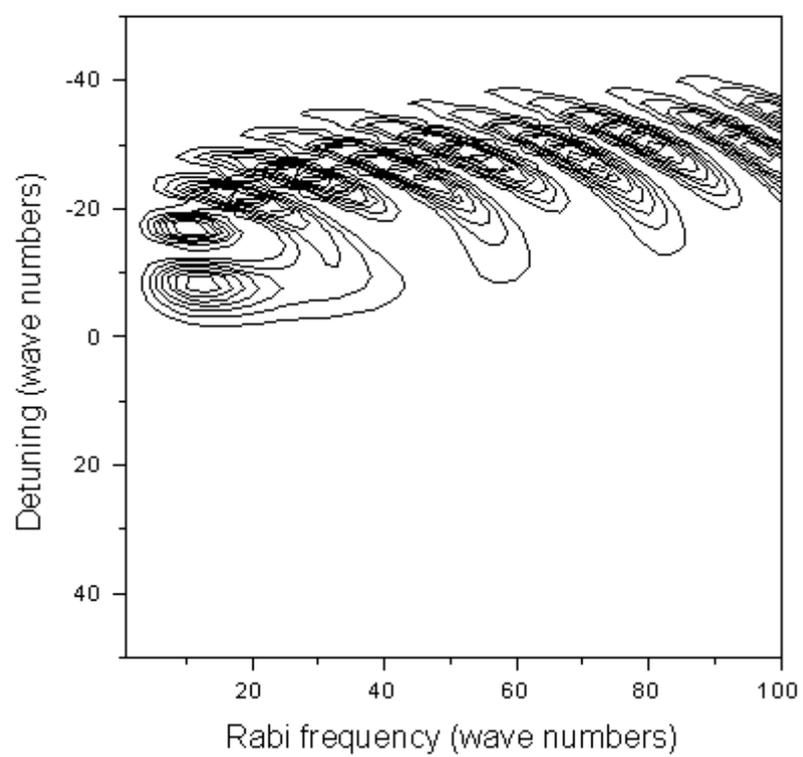
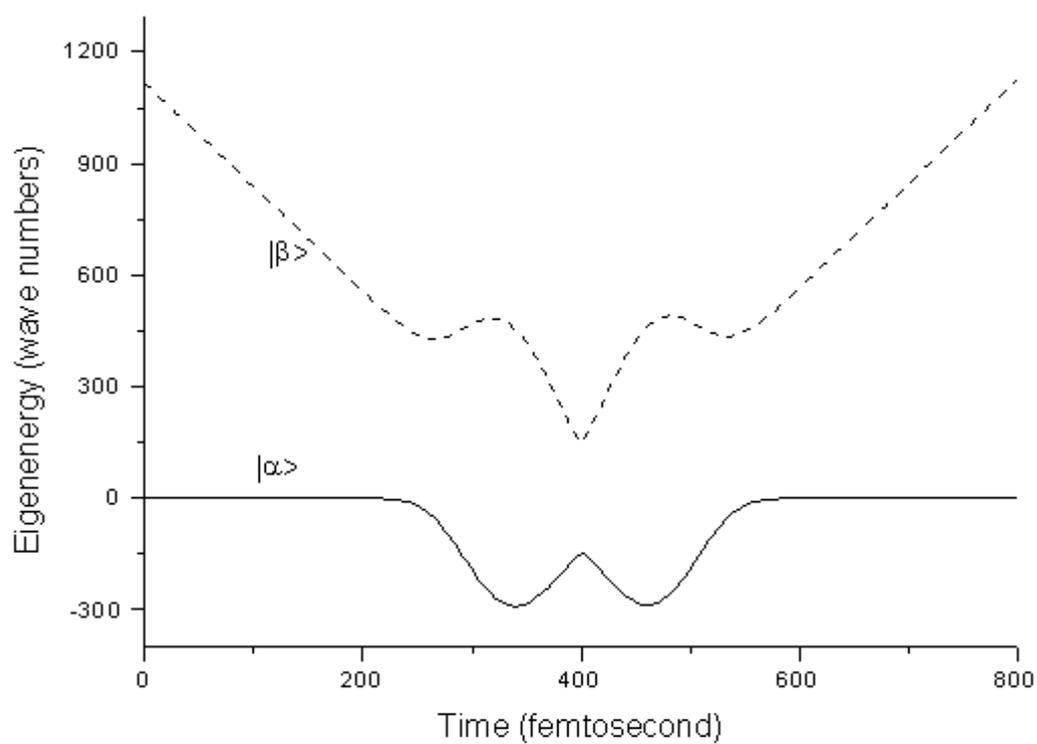
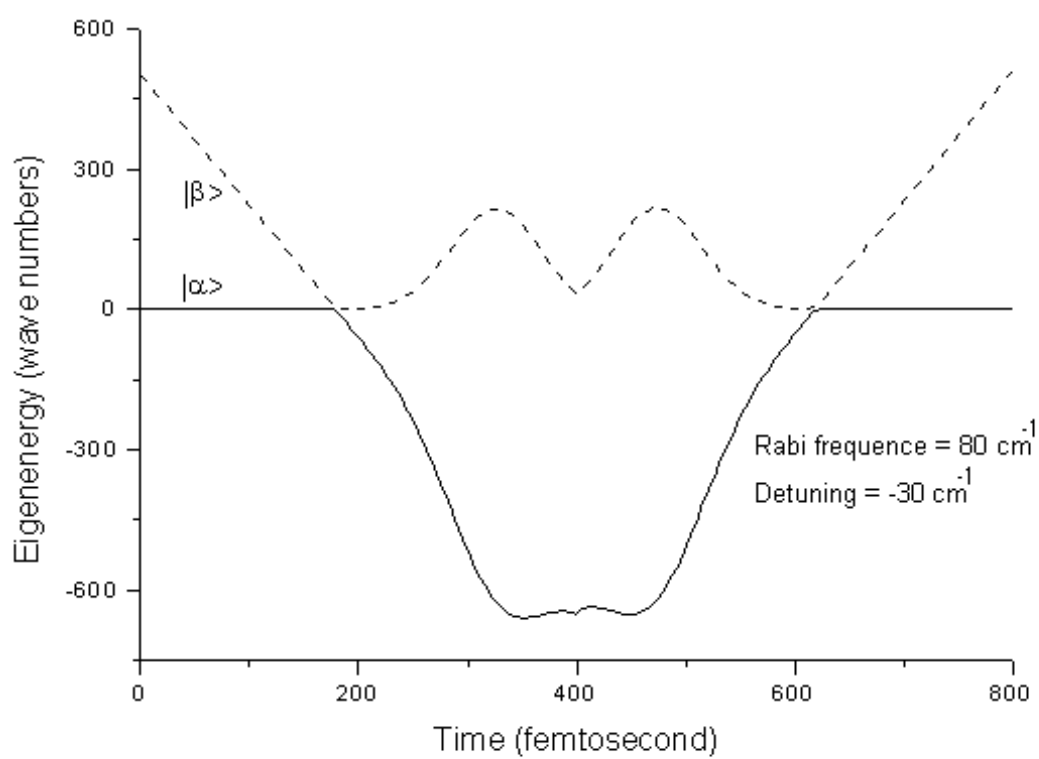
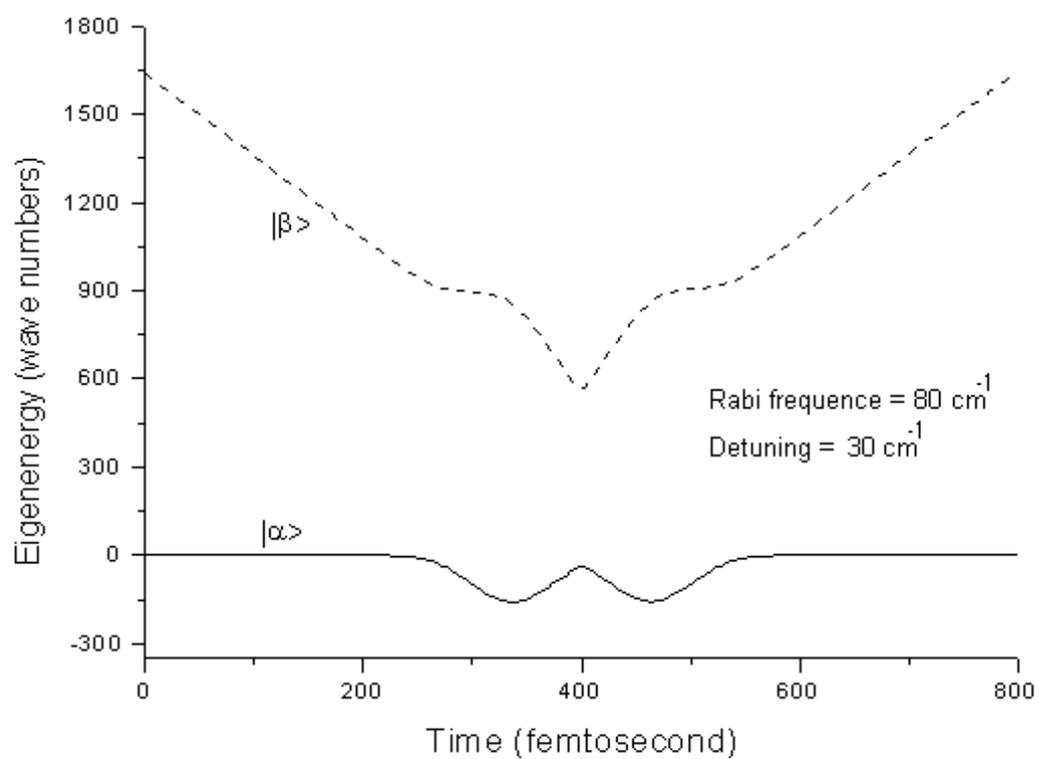
\end{document}